\documentclass[usenatbib]{mnras}
\usepackage{newtxtext,newtxmath}
\usepackage[T1]{fontenc}
\usepackage{ae,aecompl}

\usepackage{graphicx}
\usepackage{amsmath}
\usepackage{amssymb}

\title[Astrometric Confusion in the Galactic Centre]{Unrecognized Astrometric Confusion in the Galactic Centre}

\author[]{
P. M. Plewa$^{1}$\thanks{E-mail: pmplewa@mpe.mpg.de} and R. Sari$^{2}$
\\
$^{1}$Max-Planck-Institut f\"ur Extraterrestrische Physik, Garching, Germany\\
$^{2}$Racah Institute of Physics, The Hebrew University, Jerusalem, Israel\\
}

\date{Accepted 2018 February 21. Received 2018 February 21; in original form 2017 October 24}
\pubyear{2018}

\begin{document}
\label{firstpage}
\pagerange{\pageref{firstpage}--\pageref{lastpage}}
\maketitle

\begin{abstract}
The Galactic Centre is a highly crowded stellar field and frequent unrecognized events of source confusion, which involve undetected faint stars, are expected to introduce astrometric noise on a sub-mas level. This confusion noise is the main non-instrumental effect limiting the astrometric accuracy and precision of current near-infrared imaging observations and the long-term monitoring of individual stellar orbits in the vicinity of the central supermassive black hole. We self-consistently simulate the motions of the known and the yet unidentified stars to characterize this noise component and show that a likely consequence of source confusion is a bias in estimates of the stellar orbital elements, as well as the inferred mass and distance of the black hole, in particular if stars are being observed at small projected separations from it, such as the star S2 during pericentre passage. Furthermore, we investigate modeling the effect of source confusion as an additional noise component that is time-correlated, demonstrating a need for improved noise models to obtain trustworthy estimates of the parameters of interest (and their uncertainties) in future astrometric studies.
\end{abstract}


\section{Introduction}

The long-term monitoring of stellar motions in the Galactic Centre has enabled detailed studies of stellar dynamics in the extreme environment of a galactic nucleus, which benefit from the favorable conditions and growing capabilities of ground-based observations at near-infrared wavelengths \citep[for a review, see][]{2010RvMP...82.3121G}. Most notably, large proper motions of individual stars \citep{1996Natur.383..415E,1998ApJ...509..678G} and the Keplerian orbital motion of the star S2 in particular \citep{2002Natur.419..694S,2003ApJ...586L.127G} have provided convincing evidence for the existence of a central supermassive black hole identified with the compact radio source Sgr~A*. As part of the ongoing monitoring effort, more than $100$ stars in total are being tracked within the inner few arcseconds \citep[e.g.][]{2008ApJ...689.1044G,2009ApJ...692.1075G,2016ApJ...830...17B,2017ApJ...837...30G}. These stars in the immediate vicinity of the black hole were revealed as a result of several major technological advancements in high-resolution observing techniques, such as speckle imaging and the development of adaptive optics, which mitigate the effect of atmospheric turbulence, while there were parallel advancement in methods of data processing (e.g. deconvolution algorithms) and data analysis (e.g. Bayesian inference). However, even at an angular resolution close to the diffraction limit of today's largest optical telescopes, both the astrometric accuracy and precision of imaging observations of the Galactic Centre, as well as their depth, are fundamentally limited by a high level of stellar crowding.

Most importantly, stellar crowding limits the accuracy with which the extended seeing halo of the point spread function (PSF) can be determined from images. The stray halo light that is (at least partially) unaccounted for in empirical PSFs, mainly produced by a small number of luminous (WR/O-)stars, results in astrometric noise (halo noise), which represents the dominant contribution to the noise budget of all but the brightest stars in the inner nuclear star cluster \citep{2010MNRAS.401.1177F}. However, also important is the related effect of source confusion, as well as, for instance, residual image distortion, and ultimately a limited signal to noise ratio. While it is justified to exclude astrometric measurements that are firmly identified as outliers attributable to recognized confusion events involving single known stars, unrecognized confusion events involving undetected stars are by definition not separately identifiable and have to be modeled statistically.

In this paper, we present a statistical study of the astrometric effect of unrecognized source confusion in isolation, which is observationally indistinguishable from the instrumental halo noise, to better understand this effect and its impact on the monitoring of stellar motions, in anticipation of future observations using next-generation instruments, and to help further explore the limits of high-precision Galactic Centre astrometry.

\section{Methods}

\subsection{Astrometric Confusion}

If, while they are moving on the sky, two stars approach each other in projection so closely that their images overlap, a measurement of their astrometric position can be significantly biased, depending on their projected (two-dimensional) on-sky separation~${\mathbf{d}}$ and flux ratio~${L'/L}$. If one of the stars is too faint to be detected, such a confusion event is called unrecognized. However, the position of the other star will still be offset by a small amount~${\Delta\mathbf{d}}$, and it will appear slightly brighter by an amount~${\Delta L}$.

To estimate the induced changes in position and brightness in such a scenario, it is practical to minimize the total squared residuals between the true image of two overlapping sources, which are the test star and a perturbing star, and a naive model image of only a single source. This minimization can be regarded a maximum likelihood estimation if the noise in the image is Gaussian and constant over the region of interest. Both assumptions should hold for the near-infrared observations of the Galactic Centre, since they are background-dominated over any sufficiently small field of view \citep{2010MNRAS.401.1177F}. The log-likelihood function to be maximized is, up to a constant:
\begin{equation}
  \log\mathcal{L}=-\frac{1}{2}\int_{-\infty}^\infty d^2r\left(L\phi(\mathbf{r})+L'\phi(\mathbf{r}-\mathbf{d})-(L+\Delta L)\phi(\mathbf{r}-\mathbf{\Delta d})\right)^2
  \label{eq:log_L}
\end{equation}
The PSF ${\phi(\mathbf{r})}$ is assumed to be a well-sampled symmetric Gaussian function with a fixed width ${\sigma_\phi}$, which, for a first approximation, is a reasonably accurate description of the diffraction-limited core of a realistic PSF under typical observing conditions:
\begin{equation}
  \phi(\mathbf{r})=\frac{1}{\sqrt{2\pi\sigma_\phi^2}}\exp\left[-\frac{\mathbf{r}^2}{2\sigma_\phi^2}\right]
\end{equation}
The exact shape of the instantaneous PSF is variable, both spatially and over time, and may depend quite sensitively on the performance of the adaptive optics system.

\begin{figure}
  \centering
  \includegraphics[width=0.9\linewidth]{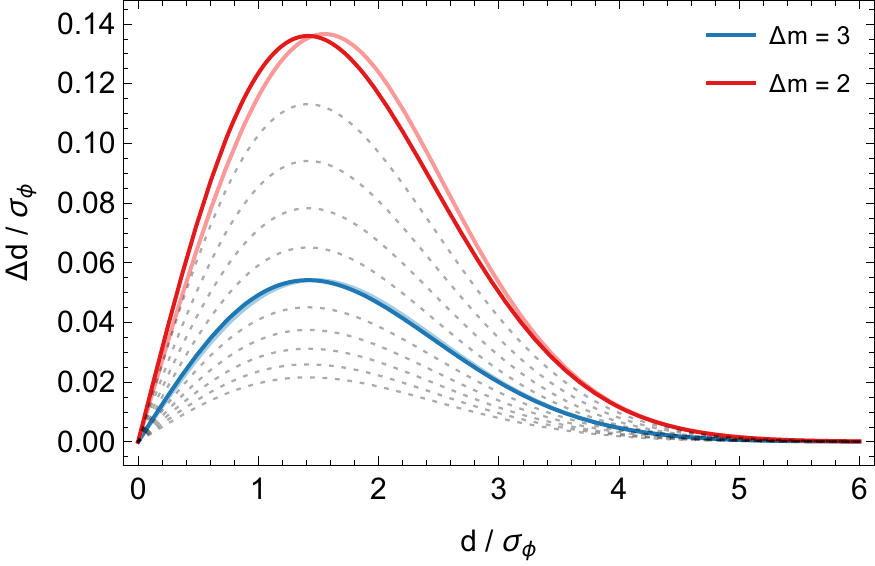}
  \caption{The astrometric offset of a test star induced by a single perturbing star, in units of the PSF width ${\sigma_\phi}$, as a function of the projected separation and magnitude difference of the two stars. Both the accurate numeric solution (to~Eq.~\ref{eq:log_L}) and the analytic approximation (Eq.~\ref{eq:delta_d}) are shown in light and dark colour, respectively, for magnitude differences of ${\Delta m=2}$ (red lines) and ${\Delta m=3}$ (blue lines). The dotted grey lines indicate how the approximate solution changes as the assumed magnitude difference varies over the range ${2\leq\Delta m\leq4}$ in steps of~$0.2$.}
  \label{fig:1}
\end{figure}

\begin{figure}
  \centering
  \includegraphics[width=0.9\linewidth]{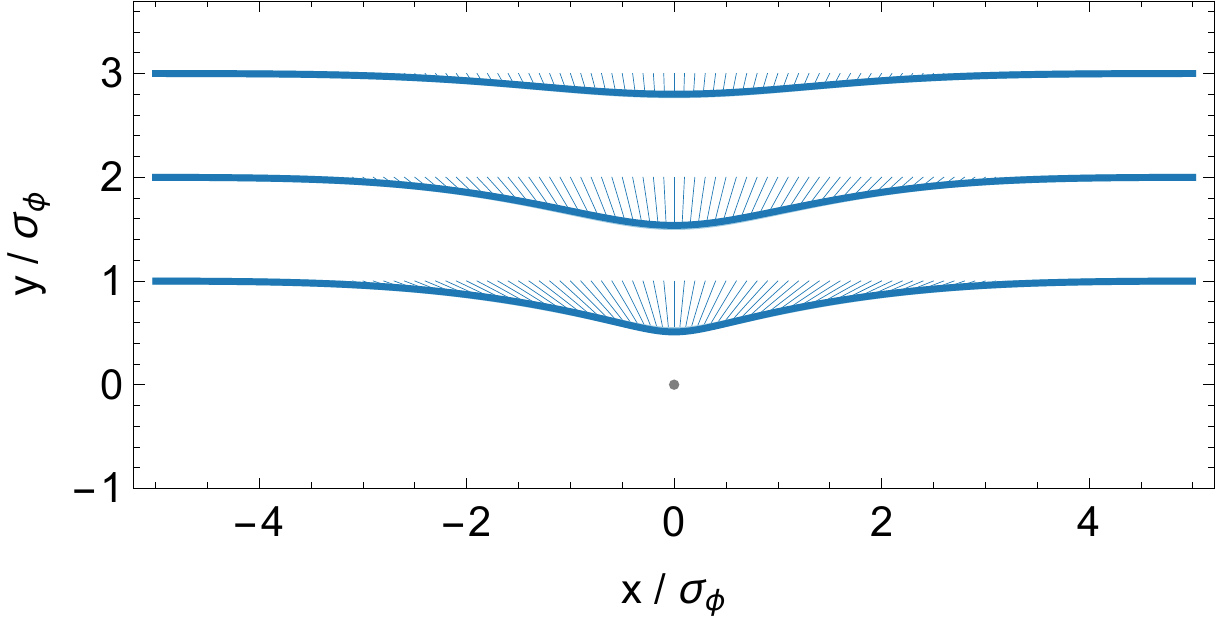}
  \caption{The two-dimensional astrometric offset of a test star as it passes a single, stationary perturbing star in projection (grey dot), for different separations at closest approach (${3\sigma_\phi}$, ${2\sigma_\phi}$ and ${1\sigma_\phi}$, where ${\sigma_\phi}$ is the width of the PSF). In all cases, the assumed magnitude difference is ${\Delta m=3}$ and the astrometric offsets are multiplied by a factor $10$ for better visibility.}
  \label{fig:2}
\end{figure}

If the perturbing star is faint enough (${L'/L\ll1}$), maximizing the likelihood function with respect to ${\Delta d}$ and ${\Delta L}$ yields:
\begin{equation}
  \frac{\Delta d}{d}\approx\frac{\Delta L}{L}\approx\frac{L'}{L}\exp\left[-\frac{d^2}{4\sigma_\phi^2}\right]
  \label{eq:delta_d}
\end{equation}
The direction of the test star's offset is towards the perturbing star, along the line connecting the two sources. This approximate solution is sufficiently good if the magnitude difference between the two stars is~${\Delta m\gtrsim2}$ (see Fig.~\ref{fig:1}), where by convention:
\begin{equation}
  \Delta m=m'-m=-\frac{5}{2}\log_{10}\left(\frac{L'}{L}\right)
\end{equation}
A perturbing star thus has the largest influence if it is located at a projected separation of ${d=\sqrt{2}\,\sigma_\phi}$ from the test star, while at separations ${d\gtrsim5\,\sigma_\phi}$ its influence becomes negligible (see Fig.~\ref{fig:1}~\&~\ref{fig:2}). In the case of more than one perturbing star, the total offset of the test star is the vector sum of the individual offsets towards each of the perturbers, and the increase in brightness is cumulative.

\subsection{The Stellar Background}
\label{subsec:background}

A large number of yet unidentified stars is expected to exist concentrated around the Galactic Centre black hole. These are sub-giant and main-sequence stars, which constitute the faint component of the existing stellar population. Despite being hidden below the detection limit of current instruments, this background (or foreground) population has an effect on the measured motions of the known, brighter stars in the central (S-)star cluster, by being a source of astrometric noise due to unrecognized source confusion.

\begin{figure}
  \centering
  \includegraphics[width=0.9\linewidth]{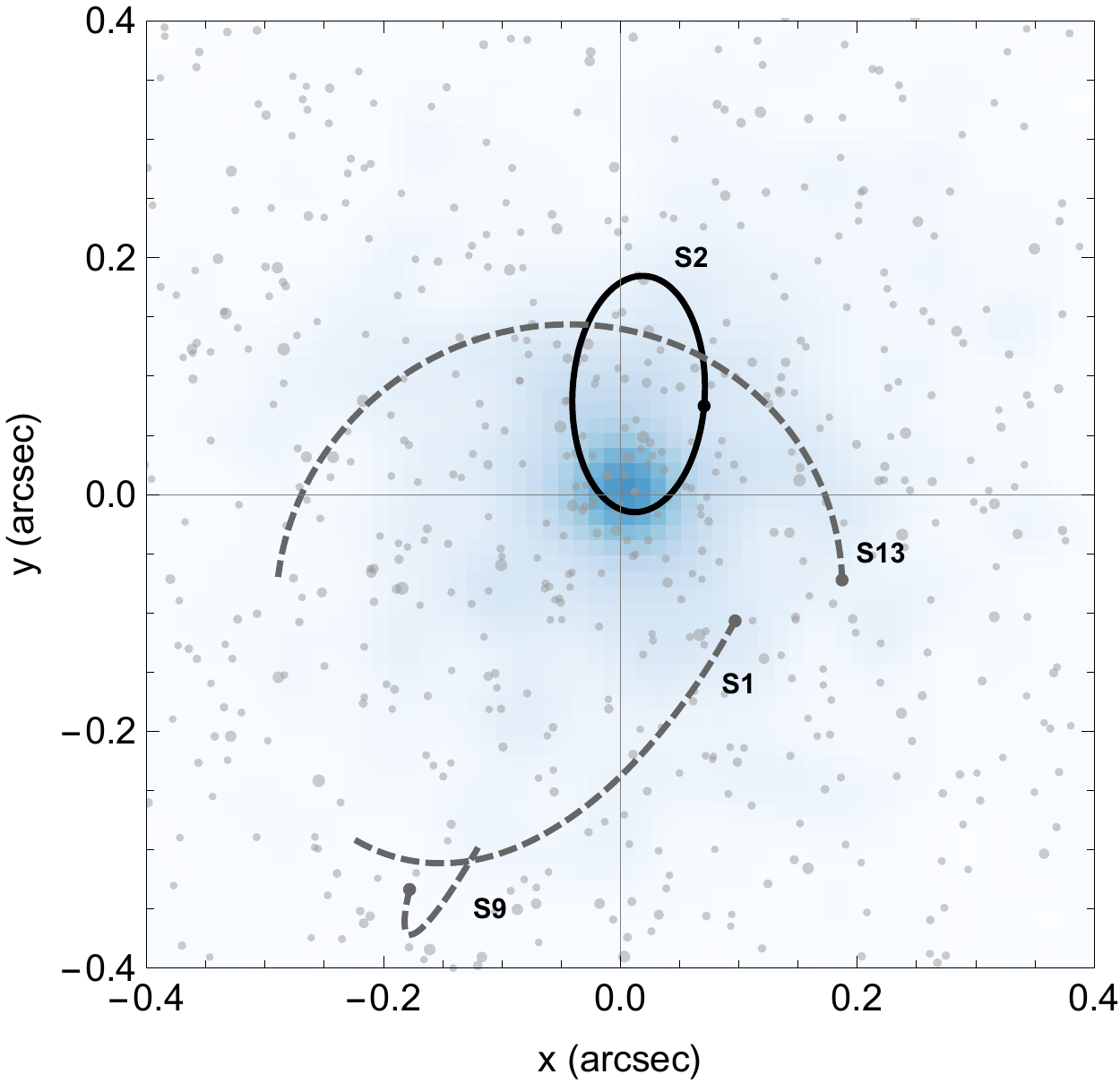}
  \caption{The on-sky trajectories of the stars S1, S2, S9, and S13 over the time of simulated observations (solid and dashed lines), starting in the year $2000$ and covering exactly one orbital period of S2 (approx. $16$ years, see \citet{2017ApJ...837...30G} for measurements of the orbital elements), overlaid on a snapshot of a sample realization of the distribution of background stars (shown as points) and their average light distribution (shown in colour).}
  \label{fig:3}
\end{figure}

We endeavor to study the properties and implications of this confusion noise by analyzing statistically the motion of a test star in an artificial field of background stars, as shown in Figure~\ref{fig:3}, while accounting for their dynamics and considering many possible realizations of their distribution. For this purpose, we assume that the stellar background population can be described by a spherically symmetric Bahcall-Wolf cusp with an isotropic velocity field \citep{1976ApJ...209..214B,Binney:2011vb}, which has a power-law distribution function ${f(\epsilon)\propto\epsilon^\gamma}$ that only depends on the stars' specific energy ${\epsilon=-v^2/2-\Phi(r)}$, where ${\Phi(r)}$ is the gravitational potential. The single free parameter ${\gamma=M_*/4M_\textrm{max}}$ depends on the stellar mass partitioning, where the stars of interest are assumed to have a given mass $M_*$ in the range ${0\lesssim M_\textrm{min}\leq M_*\leq M_\textrm{max}}$, so that ${0\lesssim\gamma\leq1/4}$. Such a Bahcall-Wolf cusp is the expected state of a dynamically relaxed stellar population surrounding a supermassive black hole. The stellar population in the Galactic Centre has a complex history \citep[e.g.][]{2011ApJ...741..108P}, is in part young and splits into at least three distinct components, when stars are grouped by their spectral types and kinematics \citep[e.g.][sec.~II]{2010RvMP...82.3121G}. We choose to examine the limiting case of a Bahcall-Wolf cusp with ${\gamma=0}$, which is analytically tractable and roughly matches the observed, slightly flatter density profile of the entire visible stellar population \citep[e.g.][]{2003ApJ...594..812G,2010ApJ...708..834B}, as well as the observed distribution of diffuse light \citep[e.g.][]{2017arXiv170103817S}, emanating from stars that could plausibly be old enough to be dynamically relaxed.

In the inner region of the nuclear cluster, stars move to good approximation like test particles in a point mass potential ${\Phi(r)=-GM_0/r}$ created by the central supermassive black hole, which has a canonical mass of ${M_0\approx4\times10^6M_\odot}$, and is located at a distance of ${R_0\approx8\,\mathrm{kpc}}$ from the Sun. To be in a stable configuration, these stars must be on bound, effectively Keplerian orbits, which can be defined by six orbital elements. The required distribution of the semi-major axes ${a=GM_0/2\epsilon}$ of the background stars, to generate a realization of a Keplerian Bahcall-Wolf cusp, is \citep{2003ApJ...596.1015S,2005PhR...419...65A}:
\begin{equation}
  p(a\mid \gamma)\propto a^{\frac{1}{2}-\gamma}
\end{equation}
The corresponding number density profile is ${n(r)\propto r^{-\frac{3}{2}-\gamma}}$. To ensure numerically that in our simulations the correct surface density profile is reproduced within the central square arcsecond, an upper cutoff of this distribution at ${a=2\,\mathrm{arcsec}}$ is sufficient. Any stars with larger semi-major axes are unlikely to be found very close to the central black hole, even in projection. We do not impose a strict lower cutoff, but exclude very few stars with a minimum (pericentre) distance ${r_p=a(1-e)<r_t}$ from the black hole, where $r_t$ is the approximate tidal radius of a typical observed B-type star in the Galactic Centre \citep[e.g.][]{2017arXiv170806353H}:
\begin{equation}
r_t\approx 1.7\,\mathrm{AU}\left(\frac{R_*}{5R_\odot}\right)\left(\frac{M_*}{10M_\odot}\right)^{-\frac{1}{3}}\left(\frac{M_0}{4\times10^6M_\odot}\right)^\frac{1}{3}
\end{equation}
The required distributions of the eccentricity~$e$, the inclination of the orbital plane~$i$, the angle of the ascending node~$\Omega$, and the argument of pericentre~$\omega$, the specific forms of which encode the assumption of isotropy for the stellar velocities and positions, are the following:
\begin{equation}
	p(e)=\begin{cases}2e & 0\leq e\leq1\\0 & \mathrm{otherwise}\end{cases}
\end{equation}
\begin{eqnarray}
	&p(\cos i)=\begin{cases}1 & 0\leq\cos i\leq1\\0 & \mathrm{otherwise}\end{cases}\\
	&p(\Omega)=\begin{cases}\frac{1}{2\pi} & 0\leq\Omega\leq2\pi\\0 & \mathrm{otherwise}\end{cases}\\
  &p(\omega)=\begin{cases}\frac{1}{2\pi} & 0\leq\omega\leq2\pi\\0 & \mathrm{otherwise}\end{cases}
\end{eqnarray}
The time of pericentre has to be sampled uniformly over the duration of one orbital period, which in turn depends on the semi-major axis:
\begin{eqnarray}
  &p(t_p\mid T(a),\,t_0)=\begin{cases}\frac{1}{T} & t_0-\frac{T}{2}\leq t_p\leq t_0+\frac{T}{2}\\ 0 & \mathrm{otherwise}\end{cases}\\
  &T(a)=2\pi\sqrt{\frac{a^3}{GM_0}}
\end{eqnarray}
The reference epoch ${t_0=2000}$ is arbitrary, but chosen to coincide roughly with the beginning of Galactic Centre observations using advanced high-resolution observing techniques, specifically adaptive optics imaging and integral field spectroscopy.

Magnitudes are assigned to the background stars such that the observed K-band luminosity function (KLF) of the nuclear star cluster is reproduced, which has a power-law slope of ${\beta\approx1/5}$ in the magnitude range ${10\lesssim m_K\lesssim18}$ \citep[e.g.][]{2003ApJ...594..812G,2010ApJ...708..834B}. We use the same slope to extrapolate beyond the detection limit to a maximum magnitude of $22$, independent of other parameters (e.g. the spatial distribution):
\begin{equation}
  p(m_K\mid\beta)\propto10^{\beta m_K}
\end{equation}
Stars are drawn randomly until a surface density of $60$ detectable stars within the central square arcsecond is reached, the magnitudes of which fall in the range ${14<m_K<m_\mathrm{lim}}$, where typically ${m_\mathrm{lim}\approx17}$. Afterwards, only the fainter stars with magnitudes larger than $m_\mathrm{lim}$ are kept, which, in the central region, are the normally undetected perturbing stars. Any stars fainter than magnitude ${m_K=22}$ have a negligible astrometric effect even on the faintest observable stars, even though they are predicted to be numerous. Higher angular resolution and sensitivity would be necessary to be able to study these fainter stars as well.

In our simulations, we do not account for any possible deviations from Keplerian orbits, for example due to a not point-like central mass distribution, encounters between stars or with remnants, or post-Newtonian effects, none of which have been detected so far \citep[e.g.][]{2012A&A...545A..70S,2017ApJ...837...30G}. These effects are either small enough to be negligible, or work on a different timescale than the effect of source confusion in the way it affects the star S2 or other fast-moving S-stars in the vicinity of Sgr~A* (see Sec.~\ref{subsec:s-stars}), and would also not significantly affect the average spatial distribution of the background stars.

\subsection{Noise Modeling \& Orbit Fitting}
\label{subsec:fitting}

We choose the star S2 (also known as S0-2) as our fiducial test star. S2 is the brightest star in the central square arcsecond and the brightest star for which an orbit has been determined. Also, the orbital period of S2 is the second-shortest one measured to date \citep[see][]{2012Sci...338...84M} and its orbit has been fully covered by observations once \citep[e.g.][]{2009ApJ...707L.114G}. For these reasons, the well-measured orbital motion of S2 provides some of the best constraints on the mass and distance of the central black hole. The orbital trajectory of S2 is shown in Figure~\ref{fig:3}, together with that of S1, S9 and S13, as well as a realization of the simulated cluster of background stars.

\begin{table}
\centering
\caption{Summary of main simulation parameters.}
\label{tab:S2}
\begin{tabular}{rcc}
  \hline
  black hole mass & $M_0$ & $4\times10^6M_\odot$ \\
  black hole distance & $R_0$ & $8\,\mathrm{kpc}$ \\
  \hline
  BW slope & $-3/2-\gamma$ & $3/2$ \\
  KLF slope & $\beta$ & $1/5$ \\
  limiting magnitude & $m_\mathrm{lim}$ & 17 \\
  PSF width & $\sigma_\phi$ & $13.8\,\mathrm{mas}$ \\
  \hline
  semi-major axis & $a$ & $0.13\,\mathrm{arcsec}$ \\
  eccentricity & $e$ & $0.88$ \\
  inclination & $i$ & $134^\circ$ \\
  longitude of the ascending node & $\Omega$ & $227^\circ$ \\
  argument of pericentre & $\omega$ & $66^\circ$ \\
  time of pericentre & $t_p$ & $2002.3$ \\
  test star magnitude & $m_K$ & $14.0$ \\
  \hline
\end{tabular}
\end{table}

As the first step in studying the background stars' effect on astrometric measurements, it is necessary to simulate observations of S2's motion, in a simplified but realistic way. Making use of the measured orbital elements \citep[Tab.~\ref{tab:S2}, see also][]{2017ApJ...837...30G}, the angular astrometric offsets $x$ and $y$ relative to Sgr~A* and the absolute radial velocity $v_z$ are evaluated at uniformly sampled times over exactly one orbital period (approx. $16$ years), to avoid any possible biases due to only fractional coverage of the orbit. The average frequency of useful imaging and spectroscopic observations is assumed to be $4$ and $3$ per year, respectively. For simplicity, neither seasonal nor technical constraints on observability are taken into account. To achieve a more uniform orbit coverage, the pericentre passage would need to be sampled at a higher rate, since S2 moves fastest during that time.

Second, the astrometric positions of the background stars are determined at the times of observations and the expected offsets due to unrecognized source confusion are added to the true positions of S2, which has a K-band magnitude of~${m_K\approx14}$ (so that~${\Delta m_K\geq3}$). The FWHM of the PSF is assumed to be a constant~${2.5\,\mathrm{px}}$ at a pixel scale of~${13\,\mathrm{mas/px}}$, which yields~${\sigma_\phi\approx13.8\,\mathrm{mas}}$ (or a FWHM of ${32.5\,\mathrm{mas}}$). This is the usual width of a restored PSF in a VLT/NACO image after it has passed the reduction pipeline \citep{2010MNRAS.401.1177F}. Since the exact dynamical configuration of the cluster of background stars is unknown, we perform separate orbit fits for many plausible, random realizations of our Bahcall-Wolf cluster model (see Sec.~\ref{subsec:background}).

Finally, additional offsets are added that are due to a generic measurement uncertainty, in the form of Gaussian noise with a standard deviation of~${\sigma=0.3\,\mathrm{mas}}$ in each of the spatial dimensions, to account mainly for halo noise but also any other sources of astrometric noise, and~${30\,\mathrm{km/s}}$ in the radial velocity dimension. For a second test case, we choose a reduced astrometric uncertainty of~${\sigma=0.1\,\mathrm{mas}}$, which approaches the signal to noise limit for an S2-like star under ideal conditions \citep{2010MNRAS.401.1177F}.

We then perform Bayesian parameter estimation on the simulated data sets, using different models to test different approaches of accounting for the effect of astrometric source confusion. In general, we use a likelihood function with three separate Gaussian components, corresponding to the two spatial dimensions and the one radial velocity dimension, which are assumed to be measured independently:
\begin{equation}
  \log\mathcal{L}=\sum_{u=x,y,v_z}-\frac{1}{2}\boldsymbol{r}_u^TK_u^{-1}\boldsymbol{r}_u-\frac{1}{2}\log\det K_u-\frac{N_u}{2}\log2\pi
\end{equation}
The residual vector~$\boldsymbol{r}_u$ (${u=x,y,v_z}$) describes the deviation of the individual measurements~$u_i$ (${i=1\dots N_u}$) from the prediction of a (Keplerian) orbit model~$f_{u,\theta}(t_i)$ evaluated at time~$t_i$, for a certain set of model parameters~$\theta$, and an additional mean function $m_{u,\theta}(t_i)$ allows for noise with non-zero mean:
\begin{equation}
  (\boldsymbol{r}_u)_i=\left(u_i-f_{u,\theta}(t_i)\right)-m_{u,\theta}(t_i)
\end{equation}
The minimum set of model parameters includes the black hole mass~$M_0$ and distance~$R_0$, the $6$ orbital elements of S2, as well as $5$ additional nuisance parameters allowing for a small drift motion of the dynamical black hole mass in the astrometric reference frame, described to first approximation by an offset in position ($x_0$, $y_0$) at the reference epoch $t_0$, a proper motion ($v_x$, $v_y$) and an offset in radial velocity ($v_z$). These additional degrees of freedom are introduced to account for the uncertainty in constructing a coordinate system that is stable over the entire time of observations \citep[see][]{2010ApJ...725..331Y,2015MNRAS.453.3234P}. The priors for $M_0$, $R_0$, the orbital parameters and the coordinate system parameters are chosen to be uniform, within reasonably broad limits.

\begin{figure}
  \centering
  \includegraphics[width=0.9\linewidth]{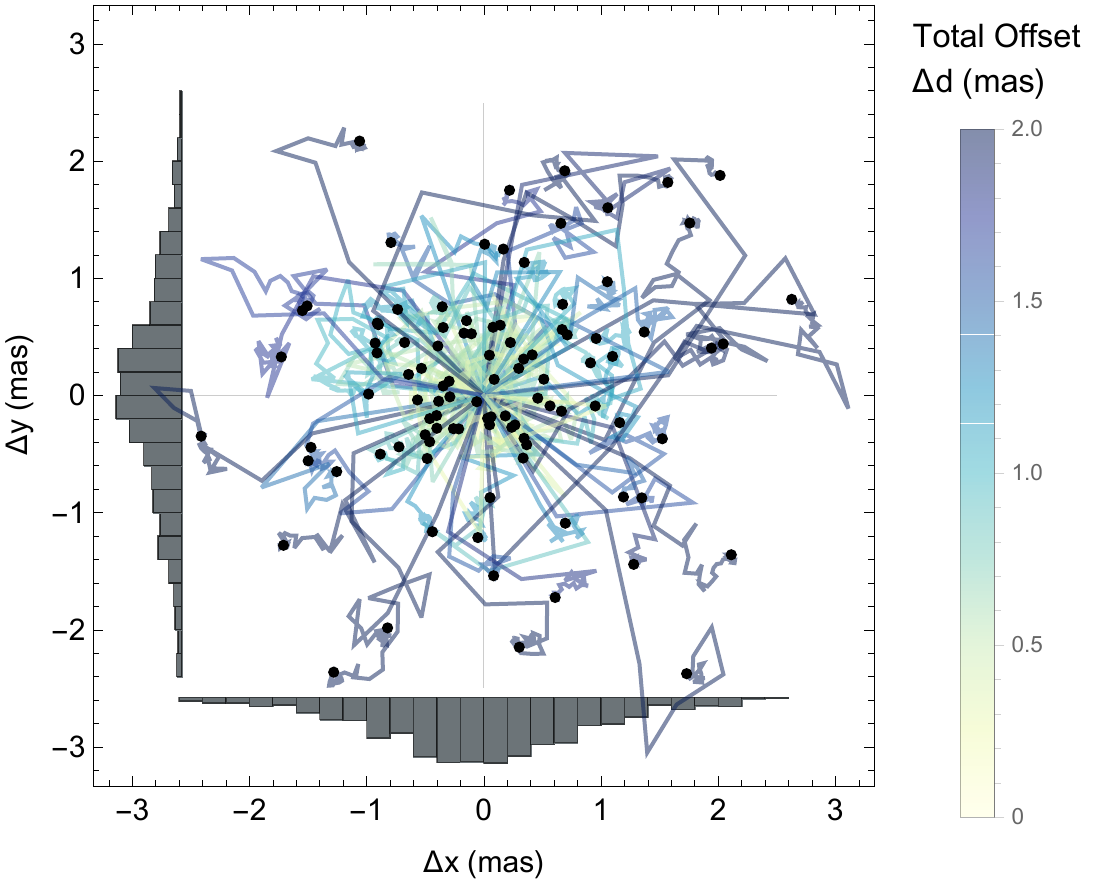}
  \caption{The total astrometric offsets of a magnitude ${m_K=15}$ test source located at the position of Sgr~A*, decomposed into successively smaller offsets caused by the individual background stars, where each line corresponds to a different realization of the background star population. Shown are 100 randomly chosen cases, together with the distributions of the total offsets along each axis, for a 10 times larger sample.}
  \label{fig:4}
\end{figure}

\begin{figure}
  \centering
  \includegraphics[width=0.9\linewidth]{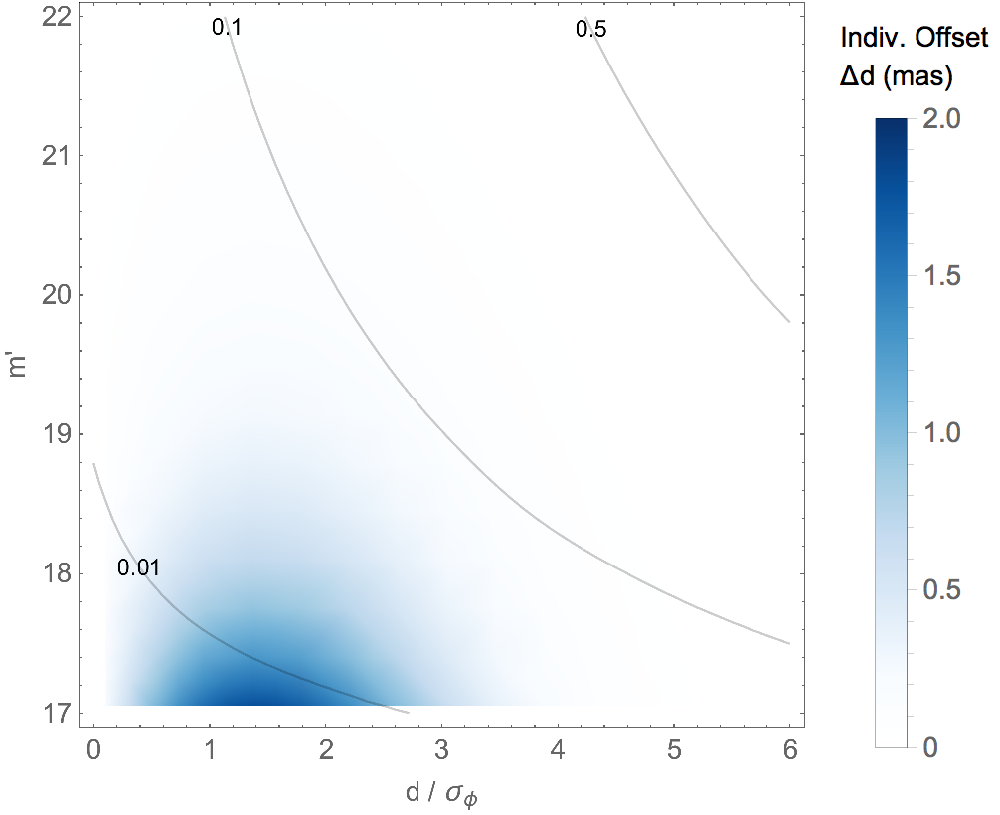}
  \caption{The size of individual astrometric offsets of a magnitude ${m_K=15}$ test source located at the position of Sgr~A* as a function of the magnitude and separation of the background stars (shown in colour), and the cumulative fraction of background stars in this parameter space (shown in contours). The effect of the brightest stars dominates (${17\lesssim m_K'\lesssim19}$), even though these stars are much less numerous than their even fainter counterparts (${19\lesssim m_K'\lesssim22}$, see also Fig.~\ref{fig:6}).}
  \label{fig:5}
\end{figure}

$K_u$ is the covariance matrix of the measurements, which is symmetric and has shape ${N_u\times N_u}$. In our standard model, $K_x$, $K_y$ and $K_{v_z}$ only contain the (true) measurement uncertainties along their diagonal as non-zero elements:
\begin{equation}
  (K_u)_{i,j}=\sigma_u^2\,\delta_{i,j}
\end{equation}
This implies or implicitly assumes that the underlying noise is known and uncorrelated over time. In reality, the true measurement uncertainties are of course unknown and have to be estimated, preferably from the data itself, if that is feasible. To account for any uncertainty in the error estimation itself and the possibility that errors may have been systematically underestimated, the likelihood function can be modified, for example by adding in quadrature another term on the diagonal of $K_x$ and $K_y$ (and possibly also $K_{v_z}$), that becomes an additional model parameter:
\begin{equation}
  (K_{x,y})_{i,j}=\left(\sigma_{x,y}^2+\eta^2\right)\,\delta_{i,j}
\end{equation}
This implies that there is some additional, intrinsic Gaussian noise with standard deviation $\eta$. However, astrometric confusion gives rise to noise that is not quite constant, but correlated over time from measurement to measurement, if only slightly. For some extended duration the test star will be confused with the same set of perturbing stars, which will bias its position in a certain direction. We attempt to model this correlated noise by making use of Gaussian processes~(GPs) and inferring three additional (hyper-)parameters instead of one, which also affect the off-diagonal elements of the covariance matrices $K_x$ and $K_y$ (yet without altering $K_{v_z}$, since the radial velocity is unaffected by astrometric confusion):
\begin{equation}
  (K_{x,y})_{i,j}=\sigma_{x,y}^2\,\delta_{i,j}+k_{x,y}(t_i, t_j)
\end{equation}
There exists a variety of valid kernel functions $k$, which can be used to encode different assumptions about noise properties \citep[see e.g.][]{Rasmussen:2006vz}. We have tested various functions from the Mat\'ern class of kernels, which are widely used because they generate noise with a certain roughness that is appropriate for describing many real-world noise processes. However, confusion events cause perturbations that are intrinsically smooth, which are therefore better described by their limiting function, an exponential squared kernel:
\begin{equation}
  k_{x,y}(\Delta t=|t_i-t_j|)=\eta^2\exp\left(-\frac{\Delta t^2}{2\tau_{x,y}^2}\right)
  \label{eq:kernel}
\end{equation}
The extra hyper-parameters are a common amplitude ${\eta=\eta_x=\eta_y}$ and two characteristic time scales $\tau_x$ and $\tau_y$. We choose to fit two independent timescales because stars on certain trajectories may encounter background stars more frequently in one direction than the other (e.g. if orbits are only partially covered by observations). The priors for these parameters are chosen to be scale-free (i.e. uniform in ${\log\eta}$ and ${\log\tau}$). The added-noise model and the GP model are fit to the same data set as the standard model, for each of many random realization of the background star population and, for a first analysis, assuming zero mean noise (${m_{u,\theta}=0}$). We also fit a control model that is identical to the standard model, except that confusion noise is not simulated, so that the assumed generic measurement uncertainty is the only noise component. In this case, all models actually yield the same result; the fitted noise amplitudes converge to zero and the time scales are unconstrained.

For each model, if we collectively denote its underlying set of assumptions as $\mathcal{M}$ (e.g. the form of the covariance matrices) and consider a simulated data set ${\mathcal{D}'=\bigcup_{u=x,y,v_z}\{u_i\}_{i=1}^{N_u}}$, the posterior probability distribution of the model parameters $\theta$ is, in the conventional notation \citep[e.g.][]{2012arXiv1205.4446H}, given by:
\begin{equation}
p(\theta\mid\mathcal{D}=\mathcal{D}',\mathcal{M})=\frac{1}{\mathcal{Z}_\mathcal{M}}\,p(\mathcal{D}\mid\theta,\mathcal{M})\vert_{\mathcal{D}=\mathcal{D}'}\,p(\theta\mid\mathcal{M})
\end{equation}
${\mathcal{L}_\mathcal{M}(\theta)\equiv p(\mathcal{D}\mid\theta,\mathcal{M})\vert_{\mathcal{D}=\mathcal{D}'}}$ is the likelihood function, defined to be the probability of obtaining the data as a function of the parameters, ${p(\theta\mid\mathcal{M})}$ is the joint prior probability distribution of the parameters, and ${\mathcal{Z}_\mathcal{M}}$ is the the fully marginalized likelihood, which is also commonly called the Bayesian evidence:
\begin{equation}
\mathcal{Z}_\mathcal{M}=\int d\theta\,\mathcal{L}_\mathcal{M}(\theta)\,p(\theta\mid\mathcal{M})=p(\mathcal{D}\mid\mathcal{M})\vert_{\mathcal{D}=\mathcal{D}'}
\label{eq:z}
\end{equation}
We obtain samples from the $13$ to $16$-dimensional posterior probability distribution using \textit{emcee}\footnote{\url{https://github.com/dfm/emcee}} \citep{2013PASP..125..306F}, a software package that implements an affine-invariant MCMC sampler \citep{Goodman:2010et}. The GP model is implemented using the package \textit{george}\footnote{\url{https://github.com/dfm/george}} \citep{2014arXiv1403.6015A}. Summary information about the posterior distribution, which is generally compact and approximately described by a multidimensional normal distribution, is stored in the form of a table of the sample means, the sample covariance matrices and several quantiles of the one-dimensional marginal distributions (accounting for the burn-in phase of the sampler). For the purpose of model comparison, the Bayesian evidence is calculated as well, using the \textit{MultiNest}\footnote{\url{https://ccpforge.cse.rl.ac.uk/gf/project/multinest/}} implementation \citep{2009MNRAS.398.1601F} of the nested sampling algorithm \citep{2004AIPC..735..395S} and its \textit{PyMultiNest}\footnote{\url{https://github.com/JohannesBuchner/PyMultiNest}} interface \citep{2014A&A...564A.125B}. For finding best-fit (i.e. maximum posterior) parameter values, we use optimization routines from the \textit{scipy}\footnote{\url{https://www.scipy.org/}} package, specifically the \textit{L-BFGS-B} algorithm.

\begin{figure}
  \centering
  \includegraphics[width=0.9\linewidth]{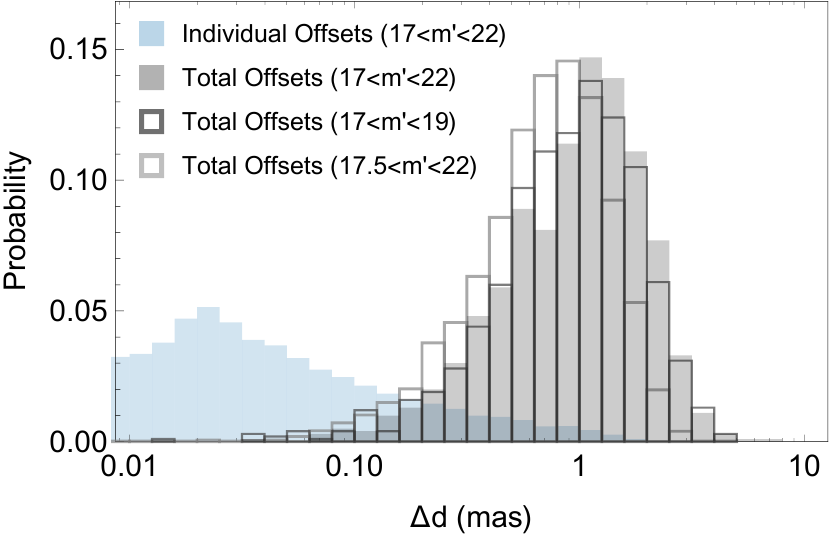}
  \caption{The probability distribution of individual astrometric offsets (shown in blue) and the total offsets (shown in grey) of a magnitude ${m_K=15}$ test source located at the position of Sgr~A*, considering many realizations of the background star population. The mode of the individual offsets is set by the maximum magnitude of the background stars ($m_\mathrm{lim}$). However, the total offset is almost solely determined by the brightest background stars, and has a typical value on the order of ${1\,\mathrm{mas}}$ in our simulations.}
  \label{fig:6}
\end{figure}

\begin{figure}
  \centering
  \includegraphics[width=0.9\linewidth]{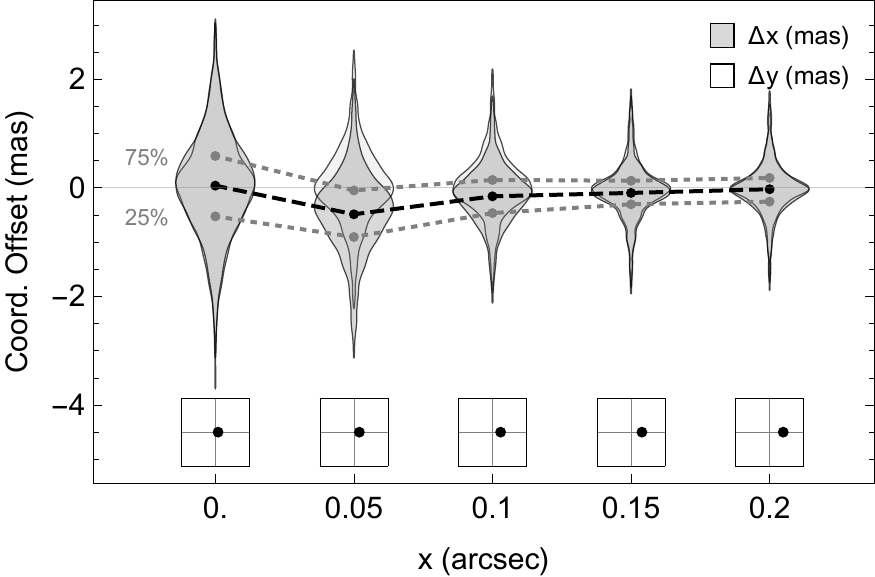}
  \caption{The distribution of total astrometric offsets of a magnitude ${m_K=15}$ test source located at different projected (on-sky) distances $x$ from Sgr~A*, where ${y=0}$, along the separation axis (${\Delta x}$) and perpendicular to it (${\Delta y}$), for many realizations of the background star population. The insets show the true position of the test source in a $1$~square arcsecond field centred on Sgr~A*. At small distances, both the amount of scatter and the overall offset towards the centre are significant compared to the usual measurement uncertainty for a source of this brightness, but both decrease notably at distances beyond ${0.1\,\mathrm{arcsec}}$, due to the decreasing density of background stars.}
  \label{fig:7}
\end{figure}

\section{Results}
\label{sec:results}

\subsection{The flares of Sgr~A*}

Without yet assuming a particular noise model, the overall amount of astrometric noise due to unrecognized source confusion in the Galactic Centre can be characterized by the effect on a test source located at the position of the black hole, at the centre of the stellar cluster, where the expected rate of confusion events is highest. We choose a magnitude $15$ source to represent a typical flare of Sgr~A*, which are known to occur sporadically \citep[e.g.][sec. VII.C]{2010RvMP...82.3121G}. These near-infrared flares are also known to show significant astrometric jitter on a mas-scale around the location of the counterpart radio source, when they are by chance captured by observations \citep{2009ApJ...692.1075G,2017ApJ...837...30G}.

The apparent astrometric offset of the test source caused by source confusion in our simulations is the superposition of many separate offsets caused by the individual background stars, which may amplify each other, but may also cancel out partially. In Figure~\ref{fig:4} we show how the total astrometric offset of the test source can be decomposed, for a number of different realizations of the background star population. The size of the induced individual offsets is also shown in Figure~\ref{fig:5} as a function of the magnitude and separation of the background stars, in comparison to their number density in this parameter space. Even though the faintest stars dominate by number, the less numerous stars with magnitudes ${17\lesssim m\lesssim19}$ at separations ${1\lesssim d/\sigma_\phi\lesssim2}$ from the test source are almost solely responsible for the total induced offset, the overall probability distribution of which is shown in Figure~\ref{fig:6}. This total offset is typically on the order of ${1\,\mathrm{mas}}$ (or ${0.07\,\sigma_\phi}$), and only rarely is it more than a factor few larger, considering again many possible realizations of the background star population. Pushing the detection limit as deep as ${m_\mathrm{lim}\approx17.5}$ would make these large, unrecognized offsets even more unlikely, but would only reduce the typical offset to ${0.7\,\mathrm{mas}}$. In any case, these numbers only describe the minimum expected astrometric scatter, since recognized confusion events involving known stars are also common at the cluster centre and would cause additional offsets.

If we place the same magnitude $15$ test source at increasing separation from Sgr~A*, now representing one of the S-stars, astrometric offsets caused by unrecognized source confusion can be expected to become smaller and less frequent, as the density of background stars decreases. However, in agreement with the study by \citet{2010MNRAS.401.1177F}, we find that at a separation of ${0.2\,\mathrm{arcsec}}$, the typical total offset of the test source is only slightly less than ${0.4\,\mathrm{mas}}$, which is still comparable to the usual measurement uncertainty for a source of this brightness. Furthermore, the offsets occur preferentially in the direction of Sgr~A*, due to the central over-density of background stars. In our test case, the average overall offset of the measured source position towards Sgr~A* is itself as large as ${0.4\,\mathrm{mas}}$ at a source separation of ${0.1\,\mathrm{arcsec}}$, as shown in Figure~\ref{fig:7}.

\subsection{The orbits of the S-Stars}
\label{subsec:s-stars}

\begin{figure}
  \centering
  \includegraphics[width=\linewidth]{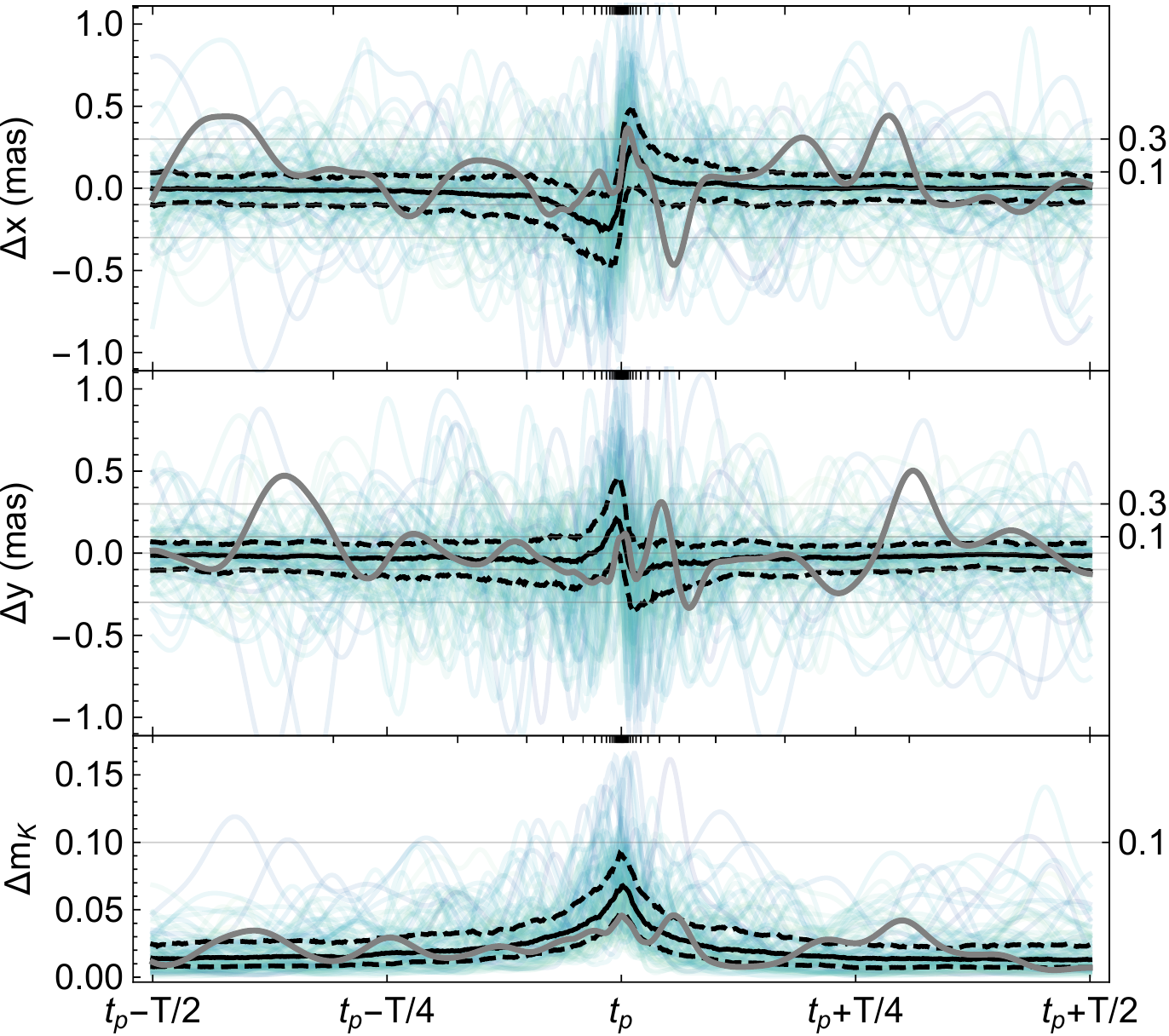}
  \caption{The expected offset in the astrometric coordinates of the star S2 due to unrecognized source confusion as it moves along its orbit (top panels), and its light curve (bottom panel). Each line corresponds to one of 100 different realization of the background star distribution, where line colour is used as a proxy for the maximum total astrometric offset (see Fig.~\ref{fig:4} for the colour scale). As an example, one randomly chosen line is highlighted in grey. The solid black lines indicate the respective median values, and the dashed lines the $25\%$ and $75\%$ quantiles. During the pericentre passage, the astrometric offset is likely to exceed the usual measurement uncertainty for a considerable amount of time, and a simultaneous increase in brightness could potentially also be detectable.}
  \label{fig:8}
\end{figure}

\begin{figure}
  \centering
  \includegraphics[width=\linewidth]{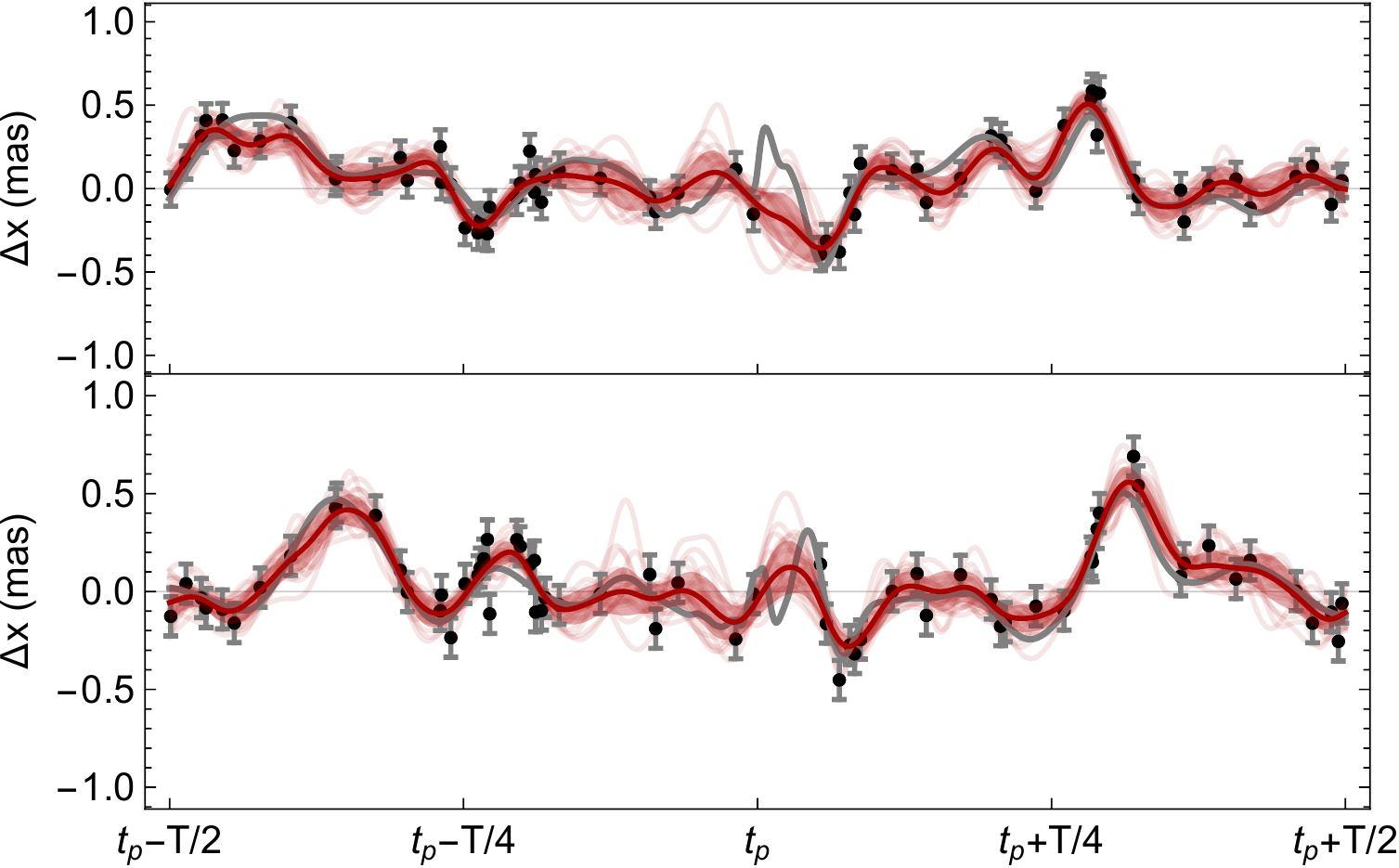}
  \caption{Results of a Gaussian process model fit to the motion of the star S2, based on a simulation corresponding to the example case highlighted in Fig.~\ref{fig:8} (grey line). The black points show the generated mock measurements as residuals with respect to the true Keplerian orbit, where the assumed measurement uncertainty is ${\sigma=0.1\,\mathrm{mas}}$, so that (unrecognized) source confusion dominates the astrometric noise budget. The perturbations caused by the confusion events are well accounted for by the noise model (red lines, same as in Fig.~\ref{fig:13}), except at pericentre, where there are too few measurements in this example.}
  \label{fig:9}
\end{figure}

The astrometric noise due to unrecognized confusion may thus contribute significantly to the noise budget of the S-stars and cause a bias in estimates of the orbital parameters, as well as the inferred mass and distance of the black hole, depending on the magnitude of each star, its on-sky trajectory and the sampling of its orbit. The resulting astrometric offsets for the star S2 are shown in Figure~\ref{fig:8} over a full orbital period, and the results of one example orbit fit are shown in more detail in Figure~\ref{fig:9}. During its pericentre passage, S2 approaches Sgr~A* closely (${R_\mathrm{min}\approx0.01\,\mathrm{arcsec}}$), due to the large eccentricity of its orbit (${e\approx0.88}$, see also Fig.~\ref{fig:3}). At this time, the probability of confusion events is highest and the astrometric offset of S2's measured position from its true position is likely to exceed the usual measurement uncertainty. The total offset at pericentre has a $50\%$ chance of being in the range from ${0.27\,\mathrm{mas}}$ to ${0.66\,\mathrm{mas}}$ and there is a remaining $7\%$ chance of an offset to occur that is greater than ${1\,\mathrm{mas}}$. These offsets can be persistent for a few months, but can also change quickly over the same time frame, since they are usually caused by only a few important confusion events, and all stars involved in these events are likely to have large proper motions.

\begin{figure}
  \centering
  \includegraphics[width=0.9\linewidth]{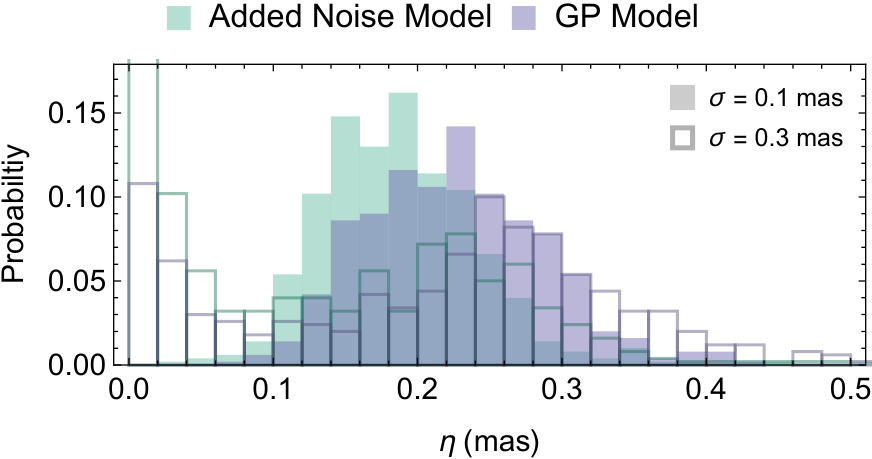}\\
  \medskip
  \includegraphics[width=0.9\linewidth]{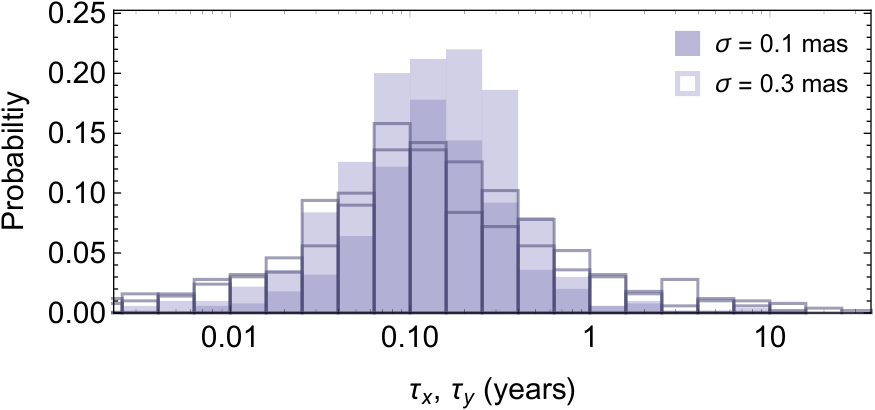}
  \caption{The mean values of the marginal posterior probability distributions for the hyper-parameters that describe the confusion noise, when fitting the orbit of the star S2 for many realizations of the background star population. Depending on the amount of measurement noise $\sigma$, the effect of confusion can sometimes be masked, resulting in small values of the noise amplitude $\eta$, but the estimated additional noise typically amounts to about ${0.2\,\mathrm{mas}}$ and is correlated on time scales $\tau_x$ and $\tau_y$ of a few ${0.1\,\mathrm{years}}$.}
  \label{fig:10}
\end{figure}

However, the pericentre passage is relatively short and S2 spends the majority of its orbital period at considerably larger distances from the black hole (${R_\mathrm{max}\approx0.19\,\mathrm{arcsec}}$). At those times, the effect of confusion can sometimes be masked by the measurement uncertainty, due to the rapidly decreasing density of background stars with increasing distance from Sgr~A*, combined with the exceptional brightness of S2. If relatively strong confusion events occur, however, they are then generally well sampled by observations. The overall confusion noise inferred using the added-noise or GP models in the simulations is around ${0.2\,\mathrm{mas}}$ on average, which is of similar order of magnitude as the typical measurement uncertainty, and is significantly correlated on a typical time scale of a few ${0.1\,\mathrm{years}}$ (see Fig.~\ref{fig:10}), which appears to be the average confusion timescale for S2.

The simultaneous brightening of S2 could be a potentially useful indicator for sufficiently strong confusion events, since it is likely to amount to a magnitude difference of a few $0.01$ magnitudes around the time of pericentre, and a change as large as $0.13$ magnitudes in cases of offsets as extreme as ${1\,\mathrm{mas}}$. The measured K-band magnitudes of S2 show a typical scatter of about $0.1$ magnitudes in the data set of \citet[][fig.~8]{2009ApJ...692.1075G}, but an improved precision may still be achieved by optimizing the image analysis for photometry instead of astrometry \citep[see e.g.][]{2010MNRAS.401.1177F,2010A&A...509A..58S}, at least for a subset of high-quality images.

\begin{figure}
  \centering
  \includegraphics[width=\linewidth]{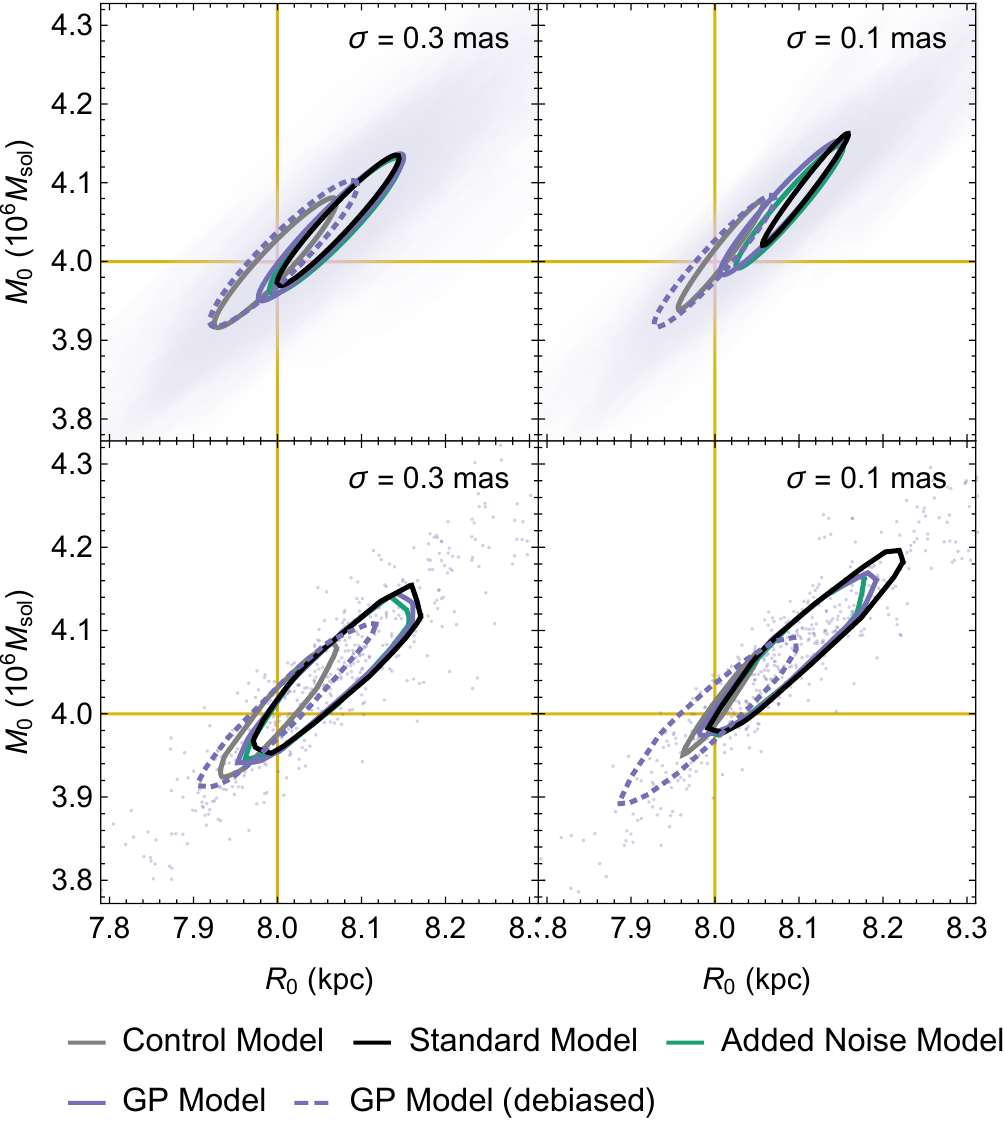}
  \caption{The typical joint posterior probability distribution of the black hole mass $M_0$ and distance $R_0$ (top panels), and the sampling distribution of the mean (bottom panels), when fitting the orbit of the star S2 for many realizations of the background star population, assuming the standard (left panels) or a reduced (right panels) measurement uncertainty $\sigma$. To highlight the effect of confusion we have assumed zero-mean noise, except in the case marked as debiased, in which the expected average offset of S2 is used as the mean function (see Fig.~\ref{fig:8}). The coloured contours indicate the 39.3\% quantile of each distribution (1-sigma) for the different models tested in our simulations, while the straight hoizontal and vertical lines indicate the assumed true parameter values.}
  \label{fig:11}
\end{figure}

As a direct result of even the unrecognized source confusion, the inferred mass and distance of the black hole can be systematically biased, and the probability of an outlier measurement based on a single star and orbit is not negligible. Both parameters are strongly correlated because the data mainly constrains the gravitational parameter $\mu$, through a measurement of the semi-major axis $a$ in angular units and the orbital period $T$:
\begin{equation}
\mu\equiv\frac{M_0}{R_0^3}=\frac{4\pi^2}{G}\frac{a^3}{T^2}
\label{eq:mu}
\end{equation}
The mass-distance degeneracy would be complete in the case of a purely astrometric data set, but it can be broken by measuring a star's radial velocity spectroscopically (in physical units) and comparing it to the star's proper motion (measured in angular units), in the context of an appropriate model (e.g. a Keplerian orbit). The typical joint posterior probability distribution of the black hole mass and distance inferred from our simulations of the motion of S2 is shown in Figure~\ref{fig:11}, as well as the distribution of the mean parameter values, considering once more many different realizations of the background star population. The properties of these distributions allow us to quantify the parameter uncertainties related to both the overall noise in the observations and our limited knowledge of the configuration of the background stars. The average bias in $M_0$ and $R_0$ is towards larger values and amounts to about~${0.06\times10^6M_\odot}$ and~${0.08\,\mathrm{kpc}}$. An estimate of the distance deviating from the true value that would, given only a measurement uncertainty of ${\sigma=0.3\,\mathrm{mas}}$, have a probability of less than~$1\%$ to occur (${\Delta R_0\gtrsim0.2\,\mathrm{kpc}}$), is predicted to be observed with a probability of about $15\%$ if source confusion is affecting the observations, but not accounted for in the noise model.

\begin{figure*}
  \centering
  \includegraphics[width=0.75\linewidth]{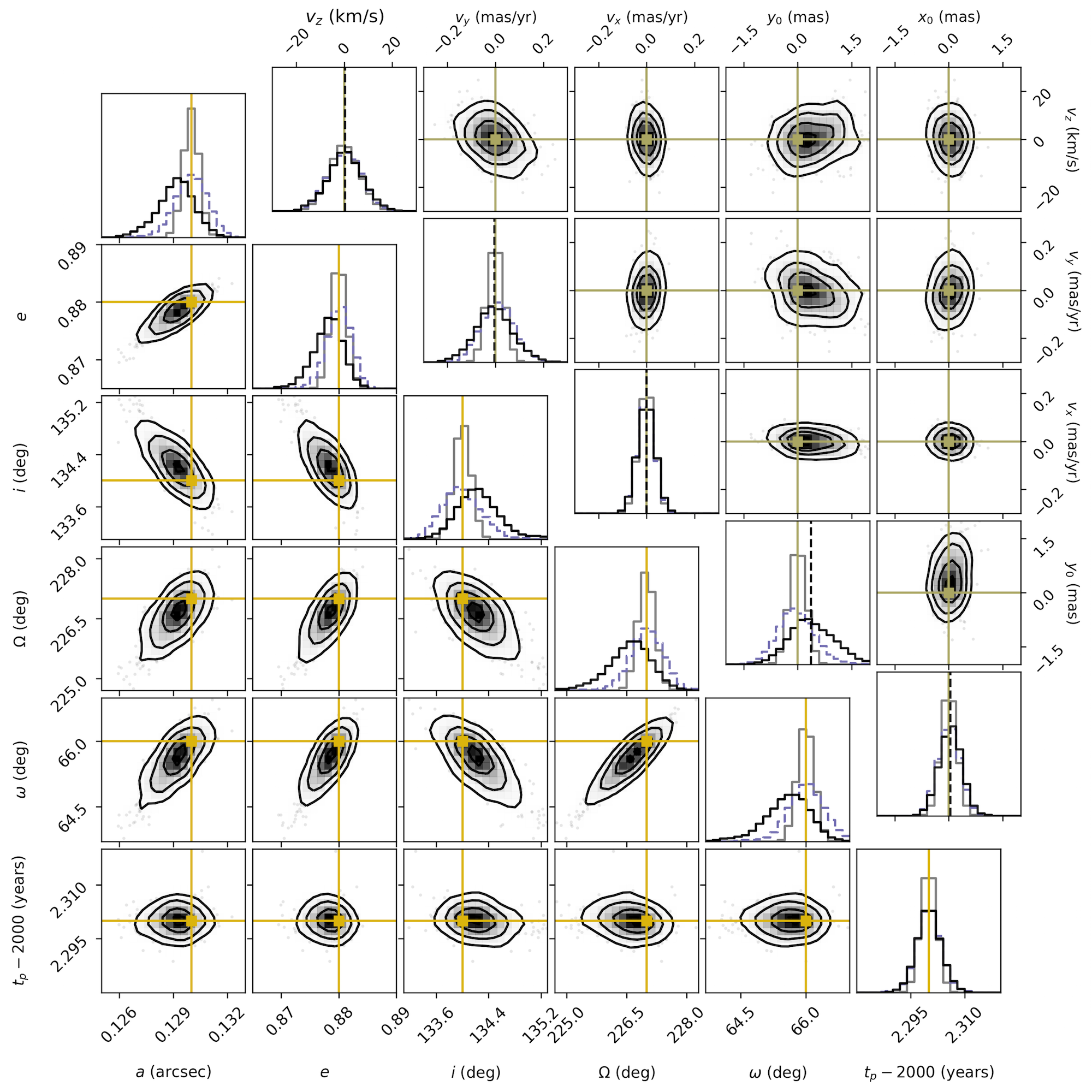}
  \caption{The mean values of the joint posterior probability distribution of S2's orbital parameters (lower left) and the coordinate system parameters (upper right) inferred from our simulations, for many possible realizations of the background star population, when using the standard orbit model to fit the motion of S2, i.e. when source confusion is not accounted for in the noise model (black lines and contours). Also shown in the one-dimensional histograms are results from the fits of the debiased GP model (dashed purple lines), as well as the control model (grey lines, same as in Fig.~\ref{fig:11}).  For all orbit fits shown here, we have assumed a generic measurement uncertainty of ${\sigma=0.1\,\mathrm{mas}}$, so that (unrecognized) source confusion dominates the astrometric noise budget. The assumed true values of the parameters are indicated by the straight horizontal and vertical lines.}
  \label{fig:12}
\end{figure*}

The jointly inferred orbital parameters are also affected by the source confusion (see Fig.~\ref{fig:12}). Perhaps most intuitively, the semi-major axis of S2's orbit is on average slightly underestimated (${\Delta a\approx-0.6\,\mathrm{mas}}$), since astrometric offsets towards Sgr~A* are somewhat more likely to occur at any time, due to the positive gradient of the surface density of background stars in that direction. This also explains the normally negative bias in the gravitational parameter ($\Delta\mu/\mu\approx-1.3\%$), since the orbital period must be unaffected (${T^2\propto a^3/\mu}$, see Eq.~\ref{eq:mu}), as is on average the time of pericentre. The eccentricity of S2's orbit is on average slightly underestimated as well (${\Delta e\approx-0.001}$) and the three orbital angles are affected similarly (${\Delta i\approx12\,\mathrm{arcmin}}$, ${\Delta\Omega\approx-12\,\mathrm{arcmin}}$, ${\Delta\omega\approx-14\,\mathrm{arcmin}}$). If an apparent shift of apocentre persists from one orbit to the next, it could complicate the prospective measurement of Schwarzschild precession, the dominant post-Newtonian effect on the orbit of S2 with an impact on astrometry. Of the coordinate system parameters, the location of the central mass is biased in the North-South direction, along which the orbit is elongated (${\Delta y_0\approx0.3\,\mathrm{mas}}$).

The average bias in all parameters is reduced by using the expected average astrometric offset of S2 as the mean function in the noise model (see Fig.~\ref{fig:8}), which changes along the orbit but has to be computed only once from simulations (assuming a certain average distribution of background stars), since the orbital trajectory can be sufficiently well determined. The added-noise and GP models yield larger parameter uncertainties as well, which would otherwise be underestimated, as would be the total noise, so that any bias is further reduced in units of the parameters' standard errors.

For the purpose of model comparison, we calculate the evidence ratios for different combinations of models (see Eq.~\ref{eq:z}), assuming that a priori ${p(\mathcal{M}_1)=p(\mathcal{M}_2)}$, so that:
\begin{equation}
\frac{\mathcal{Z}_1}{\mathcal{Z}_2}=\left.\frac{p(\mathcal{D}\mid\mathcal{M}_1)}{p(\mathcal{D}\mid\mathcal{M}_2)}\right\vert_{\mathcal{D}=\mathcal{D}'}=\frac{p(\mathcal{M}_1\mid\mathcal{D'})}{p(\mathcal{M}_2\mid\mathcal{D'})}
\end{equation}
In terms of these ratios, the GP model is favored by the simulated data in at least $90\%$ of cases, but with respect to fitting the orbit of S2 the practical difference between the GP and the added-noise model appears to be minor. However, including the additional parameters can be justified statistically, since the average evidence ratio is ${1\lesssim\log_{10}(Z_\mathrm{GP}/Z_\mathrm{add})\lesssim10}$, depending on the assumed measurement uncertainty (${0.3\,\mathrm{mas}>\sigma>0.1\,\mathrm{mas}}$). In comparison to the standard model, both the added-noise and the GP models are strongly favored.

\begin{figure}
  \centering
  \includegraphics[width=\linewidth]{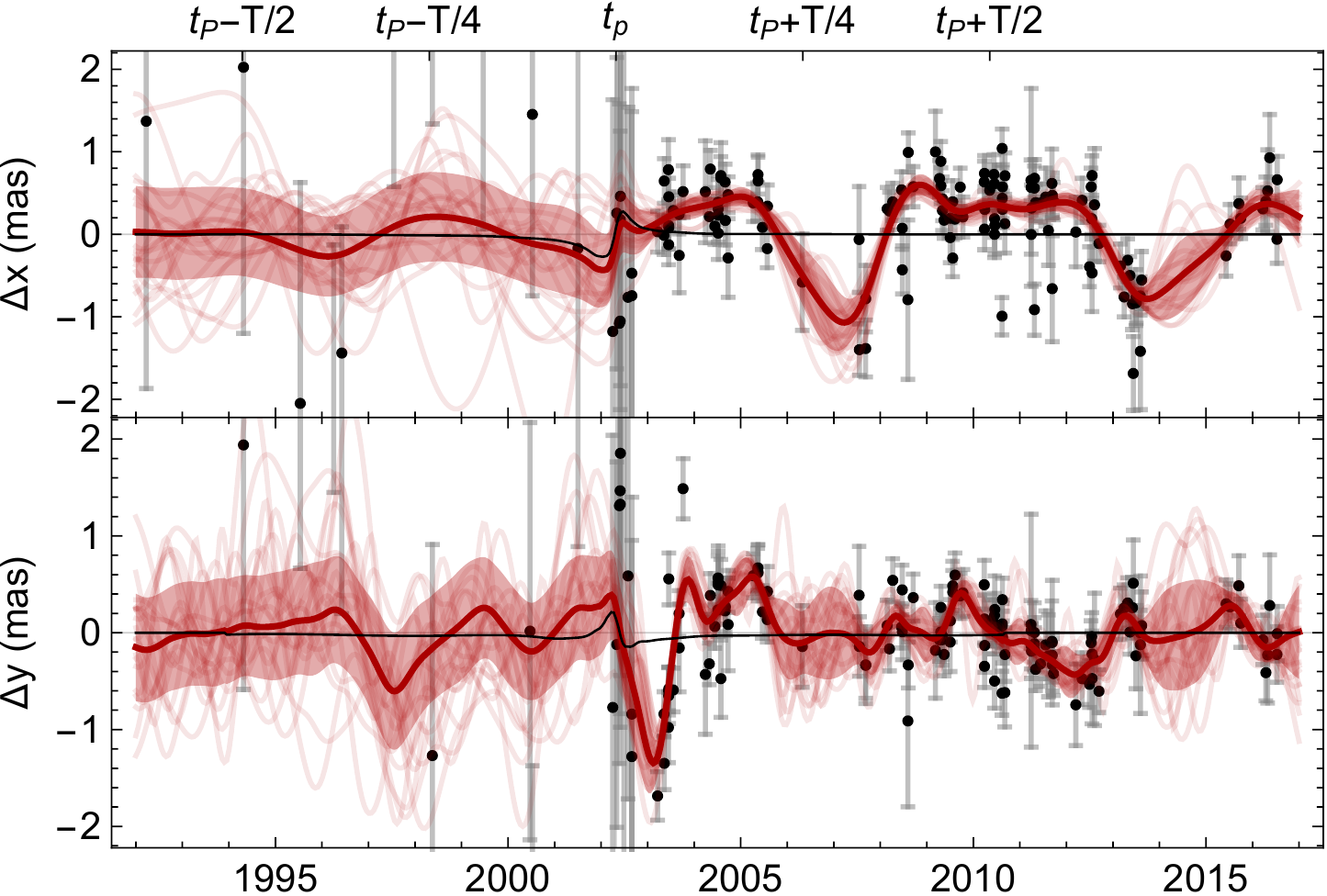}
  \caption{Results of a Gaussian process model fit to the motion of the star S2, based on the observational data set provided by \citet[][table 5]{2017ApJ...837...30G}. The black points show the astrometric residuals with respect to the best-fit Keplerian orbit. The thin red curves are randomly sampled from the posterior distribution of the inferred noise parameters, showing how the model accounts for correlations in the astrometric noise due to potential confusion events. The thick red line and error band corresponds to the posterior mean and its uncertainty (1-sigma). The thin black lines show the expected average perturbation due to unrecognized source confusion, which is included in the orbit fit.}
  \label{fig:13}
\end{figure}

Several other stars, for example S1, S9 or S13, also have well-measured orbits that constrain the gravitational potential independently, at least in principle \citep{2017ApJ...837...30G}. These stars are $0.7$ to $1.8$ magnitudes fainter than S2 and therefore potentially more affected by source confusion. However, at the moment, these stars are likely unaffected by unrecognized confusion because, in contrast to S2, they have not entered the very central region around Sgr~A* during the time over which they have been observed, but have been confined to a region where the density of background stars is much lower (see Fig.~\ref{fig:3}). The orbits of even fainter stars, for example S38, S55 or S175, are the most uncertain and usually more strongly affected by recognized confusion events than unrecognized confusion, whenever they are close to Sgr~A*.

\subsection{Application to Observations}
\label{subsec:S2fit}

As a proof of concept, we also fit our models to available observational data. The extensive S-star data set provided by \citet[][table 5]{2017ApJ...837...30G} includes $145$ astrometric and $44$ radial velocity measurements for S2, which have been collected as part of different observing programmes, but mostly at the VLT. For a detailed description of these observations and the data reduction procedure, we refer the reader to \citet{2017ApJ...837...30G} and the references therein. Prior to the orbit fitting, we have reversed the error upscaling applied by \citet{2009ApJ...692.1075G,2017ApJ...837...30G}, to retrieve the astrometric uncertainties originally estimated from the adaptive optics imaging data. We assume that any additional astrometric scatter is due to source confusion, and as such accounted for by our noise models.

The best-fit parameters are summarized in Table~\ref{tab:S2fit}, and the astrometric residuals for the fit of the (debiased) GP model are shown in Figure~\ref{fig:13}. The additional confusion noise inferred from this fit has a fairly large amplitude of ${\eta\approx0.50\,\mathrm{mas}}$ and an especially long timescale of ${\tau_x\approx0.8\,\textrm{years}}$. These values indicate that the model is accounting for astrometric perturbations due to a few pronounced, separately identifiable confusion events, rather than a series of weaker events involving undetected stars, as can be seen in Figure~\ref{fig:13}. In 2002-2003 the astrometric measurements of S2 could have likely been perturbed by S19, in 2006-2007 by S13 and in 2013 by S56, while other fainter stars and the variable infrared source Sgr~A* could have contributed as well to the overall confusion noise. The fit of the added-noise model results in a similarly large noise amplitude, yet even without explicitly accounting for the motions of any other stars (known or undetected), the GP model is able to describe the observed outlier measurements of S2 more convincingly, and is also favored in terms of the evidence ratio (${\log_{10}(Z_\mathrm{GP}/Z_\mathrm{add})\approx10}$). The inferred values for the black hole mass and distance are larger by about ${0.10-0.35\times10^6M_\odot}$ and ${0.20-0.40\,\mathrm{kpc}}$, respectively, when comparing the GP model fit to the various best-fit values of \citet[][table 1]{2017ApJ...837...30G}. This deviation is significant relative to the statistical uncertainties, and underlines the importance of the systematic effect that any kind of source confusion can have on measurements of stellar motions in a field as crowded as the Galactic Centre.

\begin{table}
\centering
\caption{Best-fit values of the added-noise and the (debiased) Gaussian process model parameters, when fitting the orbit of the star S2 using the data provided by \citet[][table 5, but see Sec.~\ref{subsec:S2fit}]{2017ApJ...837...30G}.}
\label{tab:S2fit}
\begin{tabular}{rccc}
  \hline
  & Added-Noise Model & GP Model \\
  & & (preferred) \\
  \hline
  \hline
  $M_0$ $(10^6M_\odot)$ & $4.50\pm0.18$ & $4.45\pm0.21$ \\
  $R_0$ (kpc) & $8.44\pm0.16$ & $8.53\pm0.19$ \\
  \hline
  $a$ (arcsec) & $0.1249\pm0.0009$ & $0.1228\pm0.0012$ \\
  $e$ & $0.882\pm0.002$ & $0.879\pm0.003$ \\
  $i$ & $134.8^\circ\pm0.4^\circ$ & $135.4^\circ\pm0.5^\circ$ \\
  $\Omega$ & $225.8^\circ\pm0.5^\circ$ & $226.6^\circ\pm0.7^\circ$ \\
  $\omega$ & $64.5^\circ\pm0.5^\circ$ & $64.8^\circ\pm0.7^\circ$ \\
  $t_p$ & $2002.32\pm0.01$ & $2002.32\pm0.01$ \\
  \hline
  $x_0$ (mas) & $-0.15\pm0.40$ & $0.26\pm0.74$ \\
  $y_0$ (mas) & $-1.11\pm0.55$ & $0.44\pm0.82$ \\
  $v_x$ (mas/yr) & $-0.078\pm0.041$ & $-0.084\pm0.051$ \\
  $v_y$ (mas/yr) & $-0.024\pm0.061$ & $0.050\pm0.076$ \\
  $v_z$ (km/s) & $32\pm7$ & $24\pm8$ \\
  \hline
  $\eta$ (mas) & $0.34\pm0.03$ & $0.50\pm0.13$ \\
  $\tau_x$ (yr) & & $0.78\pm0.28$ \\
  $\tau_y$ (yr) & & $0.35\pm0.08$ \\
  \hline
\end{tabular}
\end{table}

\section{Conclusions}

We have simulated long-term monitoring observations of individual stellar motions in the Galactic Centre under idealized conditions, to study in isolation the properties of astrometric noise arising from unrecognized source confusion, and specifically the ramifications for estimating the mass and distance of the central black hole by fitting stellar orbits. We emphasize that it is critical to understand the data generation process and the noise properties in particular, to be able to confidently measure any astrometric signatures of deviations from Keplerian orbits, for instance the predicted post-Newtonian deviations following the upcoming pericentre passage of the star S2 in 2018.

As the main non-instrumental source of astrometric noise, unrecognized source confusion accounts for a fundamental part of the S-stars' noise budget and can even have a significant effect on the apparent motion of a star as bright as S2. Due to the nature of source confusion in the inner nuclear star cluster, the resulting noise is temporally and spatially variable, the background stars being in constant motion themselves, yet concentrated around Sgr~A*. Exceptionally large astrometric offsets can occur in particular during the pericentre passage of S2 (or during any close enough approach of another star) and can bias the black hole mass and distance inferred from orbital motions. Recognized confusion events, as well as potential confusion with the variable infrared source Sgr~A*, add further complication to the data analysis.

The bias induced by unrecognized confusion in the black hole mass and distance can be reduced by excluding or down-weighting astrometric measurements made at and around the time of pericentre. To otherwise reduce this bias, it is necessary to account for the non-zero mean of the confusion noise, which arises from the expected central over-density of background stars. We would advocate to incorporate even a non-specific model parameterizing additional noise in some justified form, if there is a possibility that the measurement uncertainty has been underestimated. Thus estimating any additional uncertainty directly from the data will generally yield more trustworthy estimates of the parameter uncertainties, since the model for the data itself (a Keplerian orbit) is very well motivated, at least so far. This approach is also a statistically robust way to account for systematic uncertainties unrelated to confusion, for example residual image distortion. With respect to confusion specifically, a noise model based on GPs has the advantage of being able to describe time-correlated noise and is usually favored over a simpler added-noise model in our simulations, demonstrating a need for improved noise models.

Since S2 is expected to be involved in confusion events more frequently the closer it approaches Sgr~A*, it could be advantageous or even necessary to use a generalized kernel function with a varying timescale ${\tau_{x,y}(t)}$ to describe confusion noise in a GP model, e.g. Gibbs' function \citep{Rasmussen:2006vz}, if the motion of S2 is well sampled by observations during the time of closest approach:
\begin{equation}
  k_{x,y}(t_i, t_j)=\eta^2\left(\frac{2\tau_{x,y}(t_i)\,\tau_{x,y}(t_j)}{\tau_{x,y}^2(t_i)+\tau_{x,y}^2(t_j)}\right)^\frac{1}{2}\exp\left(-\frac{(t_i-t_j)^2}{\tau_{x,y}^2(t_i)+\tau_{x,y}^2(t_j)}\right)
\end{equation}
For a constant timescale ${\tau_{x,y}(t)=\tau_{x,y}}$, Gibbs' function reduces to the exponential squared kernel (Eq.~\ref{eq:kernel}), but ${\tau_{x,y}(t)}$ may be any positive function. For instance, to model a decrease in the confusion timescale around the time of pericentre, i.e. from ${t_p-\Delta t_p}$ to ${t_p+\Delta t_p}$, it would be straightforward to set
\begin{equation}
  \tau_{x,y}(t)=\tau^{(\mathrm{max})}_{x,y}+\left(\tau^{(\mathrm{min})}_{x,y}-\tau^{(\mathrm{max})}_{x,y}\right)\exp\left(-\frac{(t-t_p)^2}{2\Delta t_p^2}\right),
\end{equation}
where ${0<\tau^{(\mathrm{min})}<\tau^{(\mathrm{max})}}$, or to choose a function directly proportional to the projected separation of S2 from Sgr~A* (or some power of it). GPs may also be used to model specific, marginally recognized confusion events that apparently dominate the observed astrometric scatter of S2, so that fewer data points would have to be excluded manually (and perhaps in doubt), and to reduce any additional biases caused by these events. However, a GP model could eventually become prohibitively expensive computationally, if the number of data points will continue to grow steadily.

Although it could be possible in principle, we conclude that it is currently not feasible to deduce properties of the background star population by analyzing the properties of confusion noise, even if recognized confusion events are perfectly accounted for, unless the instrumental noise can be suppressed and all other noise processes can be thoroughly characterized. Amongst other effects, real perturbations to the orbit of S2 due to the gravitational influence of the background stars would then need to be considered as well \citep[e.g.][]{2012A&A...545A..70S}. Also, if one would want to infer the parameters of the background star population as part of the orbit fitting, using a direct forward-modeling approach, the uncertainty about the exact dynamical configuration of the background stars would make a (very inefficient) marginalization over many hyper-parameters necessary, namely the initial conditions of the background stars.

Finally, as the monitoring of stellar motions in the Galactic Centre continues, we would advocate not to rely on a single orbit of a single star for inference, but to combine measurements of many stars over many years that are affected differently by confusion and fit their orbits simultaneously, to obtain more accurate results \citep[see e.g.][]{2016ApJ...830...17B,2017ApJ...837...30G}.

\section*{Acknowledgements}

We are grateful to D.~Foreman-Mackey for helpful discussion at the 11th IMPRS Summer School on Astrostatistics and Data Mining in Heidelberg, as well as to S.~Gillessen for providing feedback on the paper draft. We also thank the anonymous reviewer for their comments, which helped to further improve the paper. Author R.~Sari is partially supported by an ICORE grant and an ISF grant.



\bibliographystyle{mnras}
\bibliography{references}



\appendix


\bsp
\label{lastpage}
\end{document}